# First-principles computation of boron-nitride-based ultrathin UV-C light emitting diodes

Jinying Wang[1*] , Kuang-Chung Wang[1], and Tillmann Kubis[1-4]

1. Network for Computational Nanotechnology, Purdue University, West Lafayette, IN 47907, USA

2. School of Electrical and Computer Engineering, Purdue University, West Lafayette, IN 47907, USA

3. Center for Predictive Materials and Devices, Purdue University, West Lafayette, IN 47907, USA

4. Purdue Institute of Inflammation, Immunology and Infectious Disease, Purdue University, West Lafayette, IN 47907, USA

* corresponding author: wang4205@purdue.edu

# Abstract

Short wavelength ultraviolet (UV-C) light deactivates DNA of any germs, including multiresistive bacteria and viruses like COVID-19. Two-dimensional (2D) material-based UV-C light emitting diodes can potentially be integrated into arbitrary surfaces to allow for shadow-free surface disinfection. In this work, we perform a series of first-principles calculations to identify the core components of ultrathin LEDs based on hexagonal boron nitride (hBN). The electrons and holes are predicted to be confined in multiple quantum wells (MQWs) by combining hBN layers with different stacking orders. Various p- and n-doping candidates for hBN are assessed, and the relative p- and n-type metal contacts with low Schottky barrier heights are identified. The findings are summarized in a concrete UV-C LED structure proposal.

# Introduction

Light-emitting diodes (LEDs) are among the most efficient electroluminescent light sources[1]. The deep ultraviolet (UV-C) light[2] and their corresponding LEDs with wavelengths shorter than 260 nm are applicable to disinfection and sterilization[3, 4], photocatalysis[5, 6], sensors[7, 8], etc[9, 10]. UV-C LEDs require materials with wide band gap, such as AlN[11] or diamond[12]. Hexagonal boron nitride (hBN) is a particularly interesting candidate[13, 14] for its reported wide band gap and two-dimensional (2D) layered structure. More importantly, it can potentially be integrated in arbitrary surfaces which can enable shadow-free surface disinfection[15-17]. The hBN-based flexible and wearable devices are imaginable as well[18-20]. Previous theoretical and experimental studies show that 2D hBN layers can be stacked in various orders[15, 21-24]. The stacking order is predicted to affect the electronic properties of hBN [15, 25] and can therefore serve as a design tool for the electronic structure of hBN-based devices. Many requirements for hBN-based LEDs have been achieved already, covering the confirmation of UV-C light emission[26, 27], the doping of hBN with p-type[28, 29] and n-type[30] carriers, the fabrication of heterostructures[14, 31], and the synthesis of large high-quality single crystals[32]. In spite of the open debate on the direct or indirect nature of the band gap for hBN[16, 17], it shows bright luminescence[16, 33] . For example, a plane-emission device based on hBN and field-emission arrays showed a stable output power of 0.2 mW at 225 nm[27].

In this work, missing elements of hBN-based UVC-LEDs are addressed. Stacking order variations are shown to confine electrons and holes in hBN-based multi-quantum wells (MQWs). This carrier confinement is found to have a direct band gap and thus allow for efficient light emission. Schottky barrier heights between hBN and various metals are calculated and preferred metal/hBN combinations are identified. The findings are summarized into a concrete UV-C LED structure proposal.

# Method

All density functional theory (DFT) calculations of this work were run with the VASP[34, 35] simulation package. In most parts, the GGA-PBE[36] functionals are applied due to their numerical efficiency. Only when explicitly mentioned, the numerically more expensive HSE06[37] functionals are used to verify key GGA results. The cutoff energy was set to 400 eV throughout. The slab model with a vacuum space of over 20 Å is used to simulate each 2D system. The Monkhorst-Pack grids are dependent on the size of the cell. For instance, a 12 × 12 × 4 (12 × 12 × 1) grid is selected in the self-consistent field iteration for bulk (few-layer) hBN. The convergence criterion for the total energy is 0.01 meV at each self-consistent field iteration. We set the Fermi level to be zero in all cases.

# Results

## 1 Stacking order and thickness effects

Five different stacking orders, AA', AB', A'B, AB, and AA (Fig.1 a-e), are known for hBN[15]. The AA' and AB stacking have been confirmed in experiments[21, 22, 24]. Although the AB' stacking is predicted to be substable, some experimental results indicate its existence[23]. The A'B and AA stacking configurations are unstable and have not been found in experiments so far[15, 38, 39]. Deterministic placement of 2D materials with van der Waals interaction between the layers has enabled atomic precision in the fabrication of modern heterostructures. It is still imaginable that deterministic placement could allow for a few-layer AA stacked hBN. Once such stacking is produced, switching from AA to other configurations require interlayer sliding forces[39, 40] similar to the friction forces of few layer graphene. This could stabilize an AA stacking configuration within a hBN heterojunction. With previous work showing marginal differences of the lattice constants of hBN with different stacking orders[15, 16, 38], an in-plane lattice constant of 0.25nm and an interplane distance of 0.333nm is adopted for all hBN systems in this work[16].

Consistent with previous theoretical results[15, 41], all hBN configurations show indirect band gaps. The band gap of bulk hBN stacked in AB, AA', AB', A'B, or AA order decreases with energies as 4.4 (5.7), 4.2 (5.6), 3.6 (4.9), 3.2 (4.4), 3.1 (4.3) eV calculated by GGA (HSE06), respectively. The hBN in stacking of AA', A'B, and AB has the top of the valence band at or near the K point and the bottom of the conduction band at the M point, while the other two stacking orders (AA and AB' stackings) have band edges at/near H and K points. For sub-stable AB' configuration, the direct conduction-to-valence-band transition is only about 0.04 eV higher than the indirect gap (Fig. S1). Although the valence and conductance bands of the five stacking forms are mainly contributed by pz orbitals of nitrogen and boron atoms, respectively, the atomic contributions and orbital arrangement are slightly different. For example, the partial charge density of the bottom of conductance band of A'B show higher contributions of pz from nitrogen atoms than the other. The conductance band edges of AA and AB' stacking forms show stronger pz-pz interlayer interaction than the systems.

There are various experimental methods that produce high-quality few-layer 2D hBN systems[24]. Few-layer hBN systems show increased band gap compared to bulk cases because of quantum confinement[42, 43]. The thickness-dependent band gap solved with GGA and HSE06 are shown in Figs.1 f-j. The band gap variation with thickness follows a second order function, i.e. it is proportional to the inverse second power of the number of layers. It is worthwhile to note that few-layer hBN systems have the same relative order of band gaps w.r.t stacking order as the bulk scenarios. The locations in the momentum space of the top of the valence band and the bottom of the conduction band of few-layer hBN agree with the respective bulk locations. For example, the 4-layer 2D hBN (Fig. 1 k-o) with AA', A'B, and AB stacking orders still show indirect band gap. The top of the valence band is near or at the K point and the conduction band minimum is at the M point. A close-to direct band gap is predicted in the 4-layer AB' and AA hBN: 4 layers of AB' hBN have the conduction band edge at the K point and the valence band edge very close to the K point. 4 layers of AA hBN yield the valence band edge at K and conduction band edges at K and M with energy differences less than 0.04 eV. The close-to direct band gaps of few-layer AA and AB' stackings are the most relevant hBN candidates for LEDs. Mengle et. al[41] have also noticed that the bilayer AA and AB' stacking hBN show direct gap. Furthermore, they constructed a randomly stacked 10-layer hBN and found it has a direct gap which is 49 meV smaller than the indirect gap. To be noticed, both GGA and HSE06 functionals give similar band curves, charge density, and relative gap orders, and thus GGA can serve as numerically efficient estimate when specific gap values are not a critical simulation parameter.

## 2 Multiple quantum wells (MQWs)

Light emitting diodes typically use quantum wells to trap electrons and holes for a higher radiative recombination probability[9, 44]. The stacking order dependence of the effective band gap of hBN layers allows designing all-hBN multi-quantum well heterostructures. The AA' stacking direction yields a large band gap (see Fig.1) and can serve as barrier material. Embedding either AA or A'B stacked hBN layers in AA' layers is expected to form quantum wells for electrons and holes. This is confirmed in Figs.2, which show two periodic superlattices of hBN composed of several hBN layers of varying stacking orders. Fig. 2a shows a superlattice with 4 layers of AA' stacking, alternating with 5 layers of AB' stacking (labeled MQW-T1). Note that the interface between AB' and AA' effectively represents an AA bilayer. Another superlattice composed of unit cells of 3 layers of AA stacking and 6 layers of AA' stacking is displayed in Fig. 2b (labeled MQW-T2). The MQW-T1 structure yields a direct band gap of 3.50 (4.79) eV in GGA (HSE06) calculations. By analyzing the partial charge density, we find clear quantum confinement of the states in the AA layers (Fig. 2e). The confined states are in the energy range between -0.21 (-0.24) eV to 0.00 (0.00) eV and in AB' region for energy of 3.50 (4.79) eV to 4.16 (5.50) eV solved with GGA (HSE06). The energy range of the confined states equals the band offsets between wells and barriers in Fig. 2c. For MQW-T2, the band gap is 3.39 (4.60) eV in GGA (HSE06) calculations, and the conduction band edges at K and M points differ by 0.01 (0.07) eV. Similar state confinement in wells has been found with the band offsets of 0.56 (0.63) eV and 0.38 (0.43) eV for electron and hole, respectively. The band offsets of the proposed MQWs are comparable to other UVC LEDs[10]. Bulk hBN has shown considerable photoemission probability[33]. However, the light emission is expected to be greatly enhanced by breaking the stacking symmetry[41] and tailoring the stacking sequences carefully, especially introducing AA or AB’ stacking wells, due to the direct (quasi-direct) band gap and carrier confinement. It is worthwhile to mention that bulk or few-layer hBN without any AA or AB’ stacking sequences always have indirect gap according to our tests.

# 3 Doping of hBN

LED structures are typically built upon a pn junction. To facilitate the doping profile in experiments, foreign atoms are usually implemented in the respective material[28, 30, 45-47]. We model the doped hBN with a superlattice of 4x4 primitive hBN unit cells and replace one B atom with either Si, Mg, or Be. The relaxation of these doped structures results in an increase of the unit cell by about 2 % (shown in Figs. 3 a-c). The predicted electronic density of states (DOS) in Fig. 3 d shows that doping of a monolayer of hBN with Si atoms moves the Fermi level by about 0.2 eV into the conduction band. In contrast, the Fermi level lies 0.3 eV deep in the valence band for Be and 0.4 eV for Mg doping. Note that Si and Be atoms have been shown in experiments to yield n-doping and p-doping in hBN, respectively[28, 30].

# 4 Schottky barrier of hBN-metal contacts

Schottky barrier heights are dependent on the chemical potential difference and atomic configurations of the respective contacts. Every different doping setup has a different appropriate contact material. The Schottky barrier resulting from four lead materials, i.e. Co, Ni, Au, and graphite when in contact with hBN is assessed here. Following Ref. [48], hBN/metal contacts are modeled with 4 layers of AA’ hBN set on 6 layers of the metal contacts. The cross sectional size of the sample is 1x1 unit cells for hBN on hpc-Co (0001), fcc-Ni(111), and hpc-graphite (0001). When hBN is interfaced with fcc-Au(111), 2x2 unit cells are used (see Figs. 4 a-d). The relaxation of two layers of hBN and two layers of the respective contact material closest to the interface is based on the optB88-vdW-DF functional which includes the van der Waals interactions between hBN and the contact material. The remaining layers are assumed to be unstrained. The resulting lattice mismatch is less than 2% in all considered interfaces. Electronic band structures solved with GGA-PBE functionals on the relaxed structures are shown in Figs. 4e-h. The projections on B and N atoms are highlighted in green and red, respectively. The comparison with the total band structure (black lines in Figs 4e-h) allows to identify the electron and hole Schottky barriers ($\Phi_e$ and $\Phi_h$) as the difference of the hBN/metal contact Fermi levels with the respective hBN band edges

$$\begin{aligned} \Phi_\mathrm{e} &= E_\mathrm{C} - E_\mathrm{F} \\ \Phi_\mathrm{h} &= E_\mathrm{F} - E_\mathrm{V} \end{aligned} \quad (1),$$

respectively. Here $E_F$ is the Fermi energy level of the hBN/metal contact, $E_C$ ($E_V$) is the conduction (valence) band edge of intrinsic hBN, which can be extracted by projected atomic contributions of boron and nitride atoms to the total band structure of hBN/metal interfaces. The results in Table 1 identify Au and graphite as good contact metal candidates for p-doped hBN. Co and Ni are suitable to connect n-doped hBN. Note that the Schottky barriers for hBN/fcc-Co(111) show very similar heights as hBN/hcp-Co(0001) (within 0.1 eV, see Fig. S2). When hBN is doped, the Schottky barriers get further reduced (as exemplified in Fig. (S3)).

## 5 Summary

Figure 5 summarizes the findings of our DFT calculations into a concrete proposal of an ultrathin UVC-LED structure. The MQWs with clear quantum confinement are built up using hBN with different thickness and stacking orders. Au and graphite are good candidates for p metals, while Ni and Co can be used for n metals. The n-doping and p-doping hBN can be realized by implanting Si and Be/Mg atoms, respectively. Our results give clear guidance to experimentalists and is possible to be realized soon.

# Acknowledgements

The authors acknowledge the Texas Advanced Computing Center (TACC) at The University of Texas at Austin for providing HPC resources that have contributed to the research results reported within this paper. This research used resources of the Oak Ridge Leadership Computing Facility, which is a DOE Office of Science User Facility supported under Contract DE-AC05-00OR22725. The work was supported by NSF EFRI-1433510. We also acknowledge the Rosen Center for Advanced Computing at Purdue University for the use of their computing resources and technical support. T.K. acknowledges support from Silvaco. We are grateful to helpful discussion with James Charles and Yuancheng Chu.

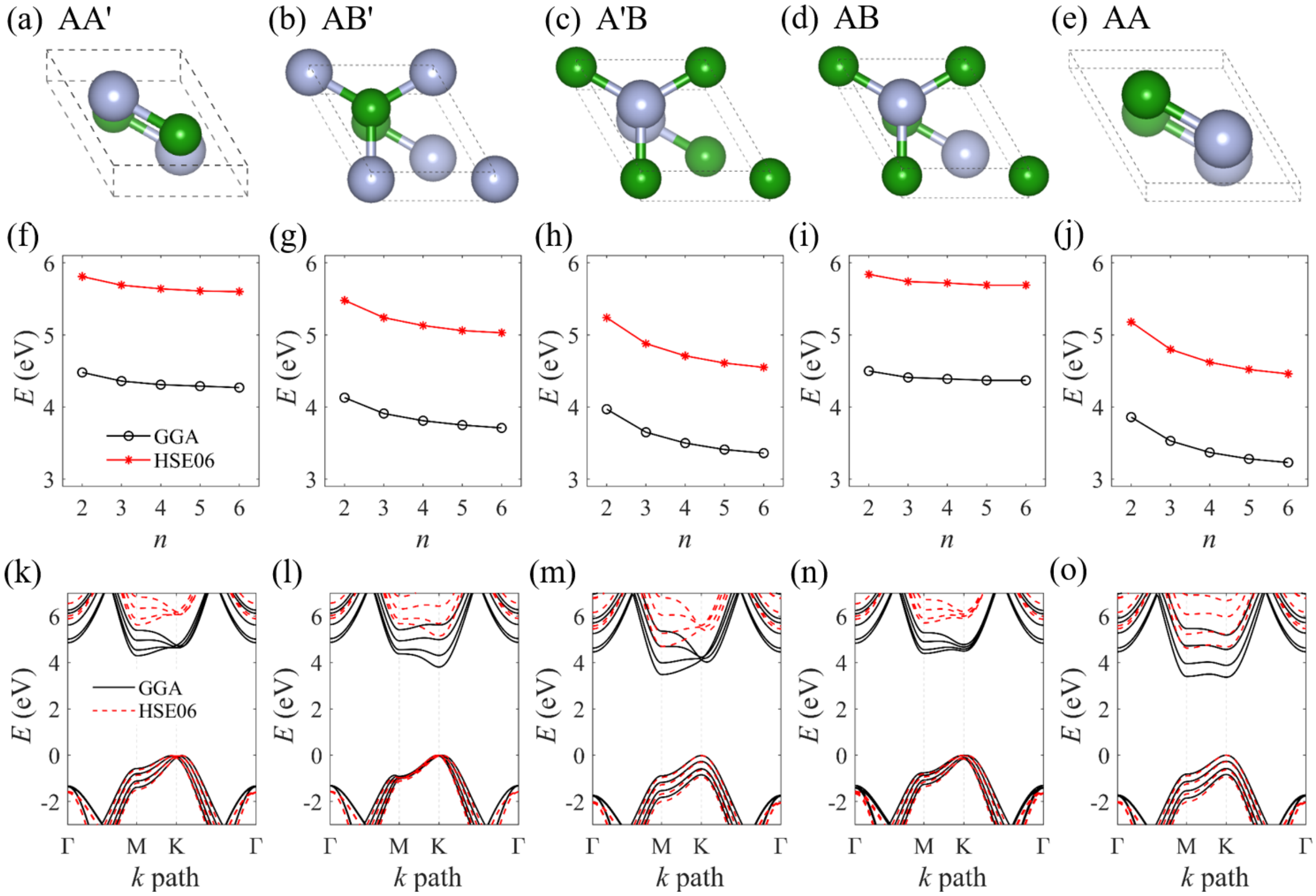

Figure 1. Atomic structures of bulk hBN with (a) AA', (b) AB', (c) A'B, (d) AB, and (e) AA stacking orders. Dependence of bandgap of few-layer 2D (f) AA', (g) AB', (h) A'B, (i) AB, and

(j) AA stacking hBN on the number of layers. Band structure of 2D (k) AA', (l) AB', (m) A'B, (n) AB, and (o) AA stacking hBN with 4 layers.

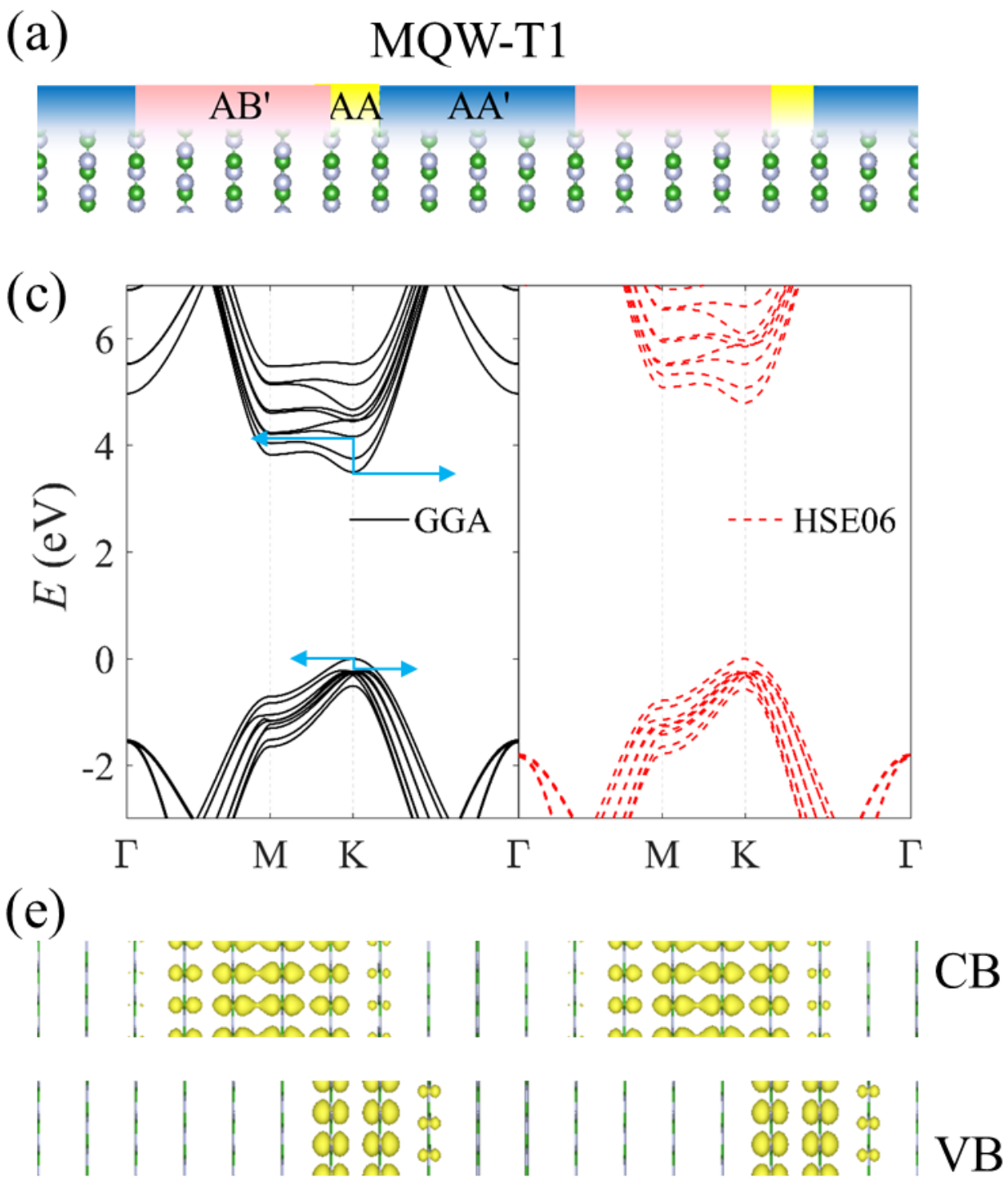

Figure 2. Atomic structures of two multiple quantum well superlattices: (a) MQW-T1 and (b) MQW-T2. Their band structures (c,d) are solved with GGA-PBE and HSE06. The partial charge density in the energy range of the barrier/well band offsets (marked by blue arrows in c and d) for electrons and holes are shown in (e, f), respectively. The isosurfaces in (e,f) show the partial charge density at 0.001 e/Å$^{-3}$.

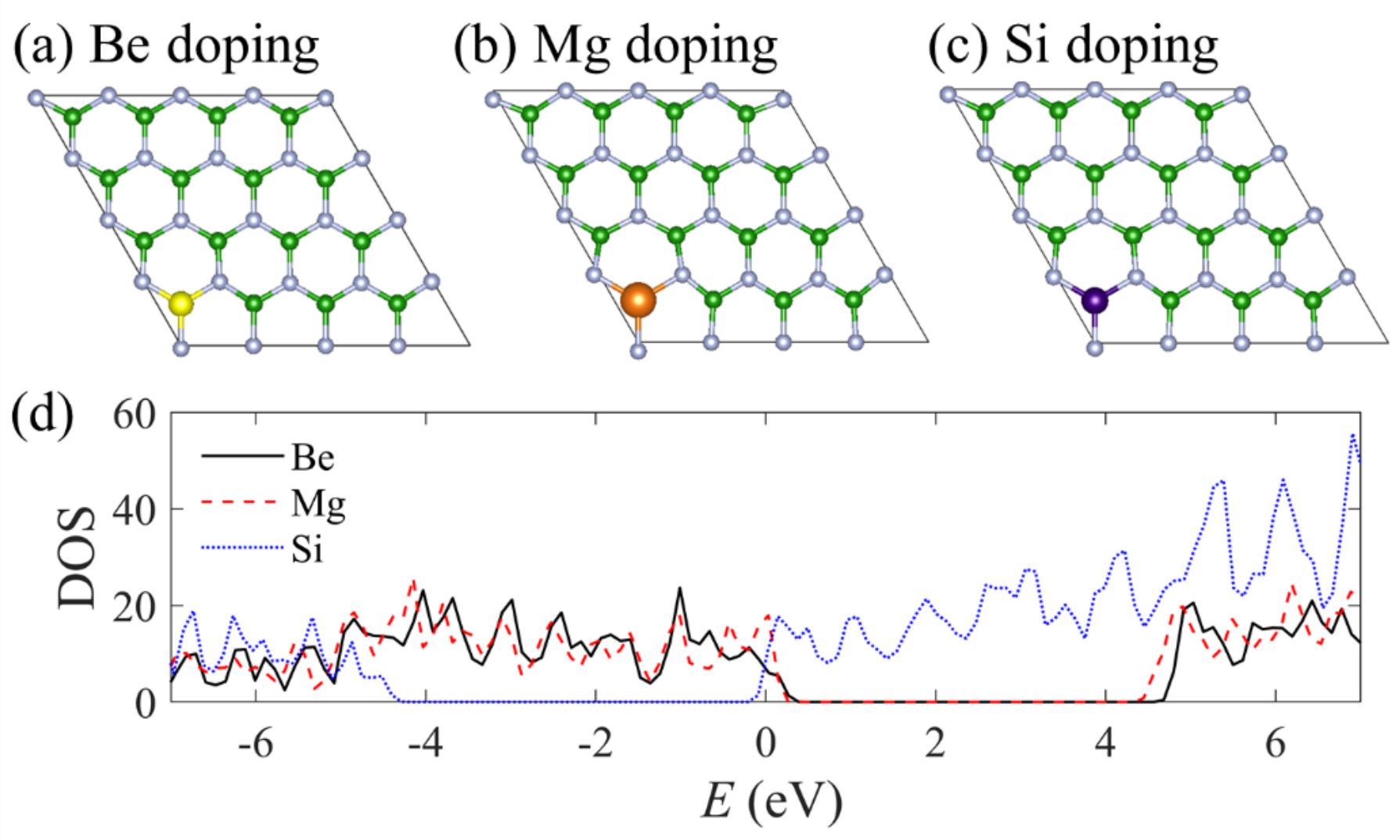

Figure 3. Atomic structures of (a) Be-doped (b) Mg-doped (c) Si-doped hBN and (d) their DOS calculated by GGA-PBE.

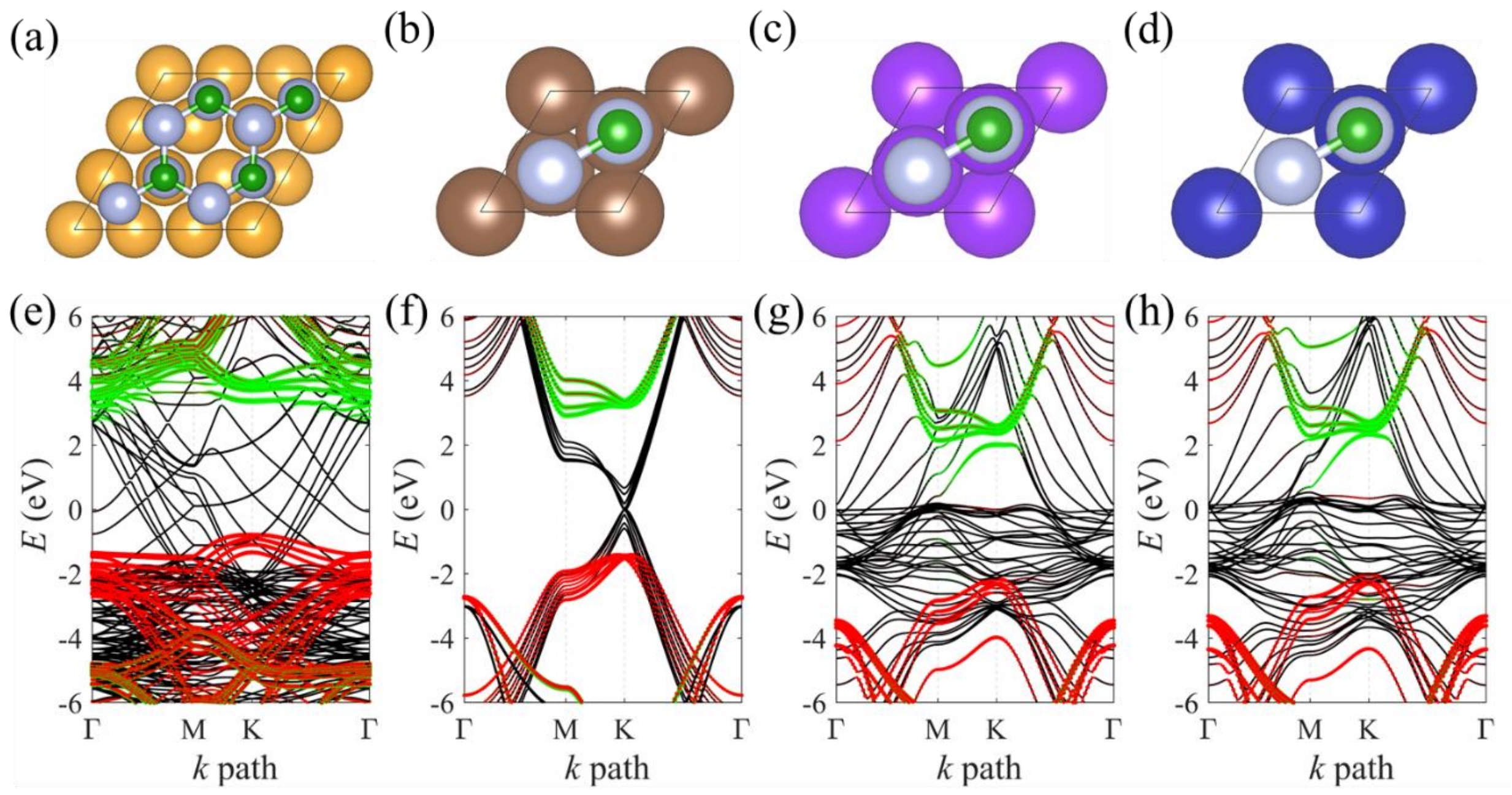

Figure 4. Atomic configuration and band structures of 4L-hBN on (a) Au (111), (b) graphite (0001), (c) Ni (111), and (d) Co (0001). The full bands are solid lines in black, projections on B atoms are marked in green and projections on N atoms in red.

Table 1 The electron and hole Schottky barrier heights ($\varphi_e$ and $\phi_h$ for contacts between intrinsic hBN and several metals (Au, graphite, Ni, Co).

| | | Au | C | Ni | Co |
|---|---|---|---|---|---|
| 4L-hBN | $\phi_p$ (eV) | 0.79 | 1.42 | 2.17 | 2.33 |
| | $\phi_n$ (eV) | 3.36 | 2.91 | 2.01 | 2.03 |

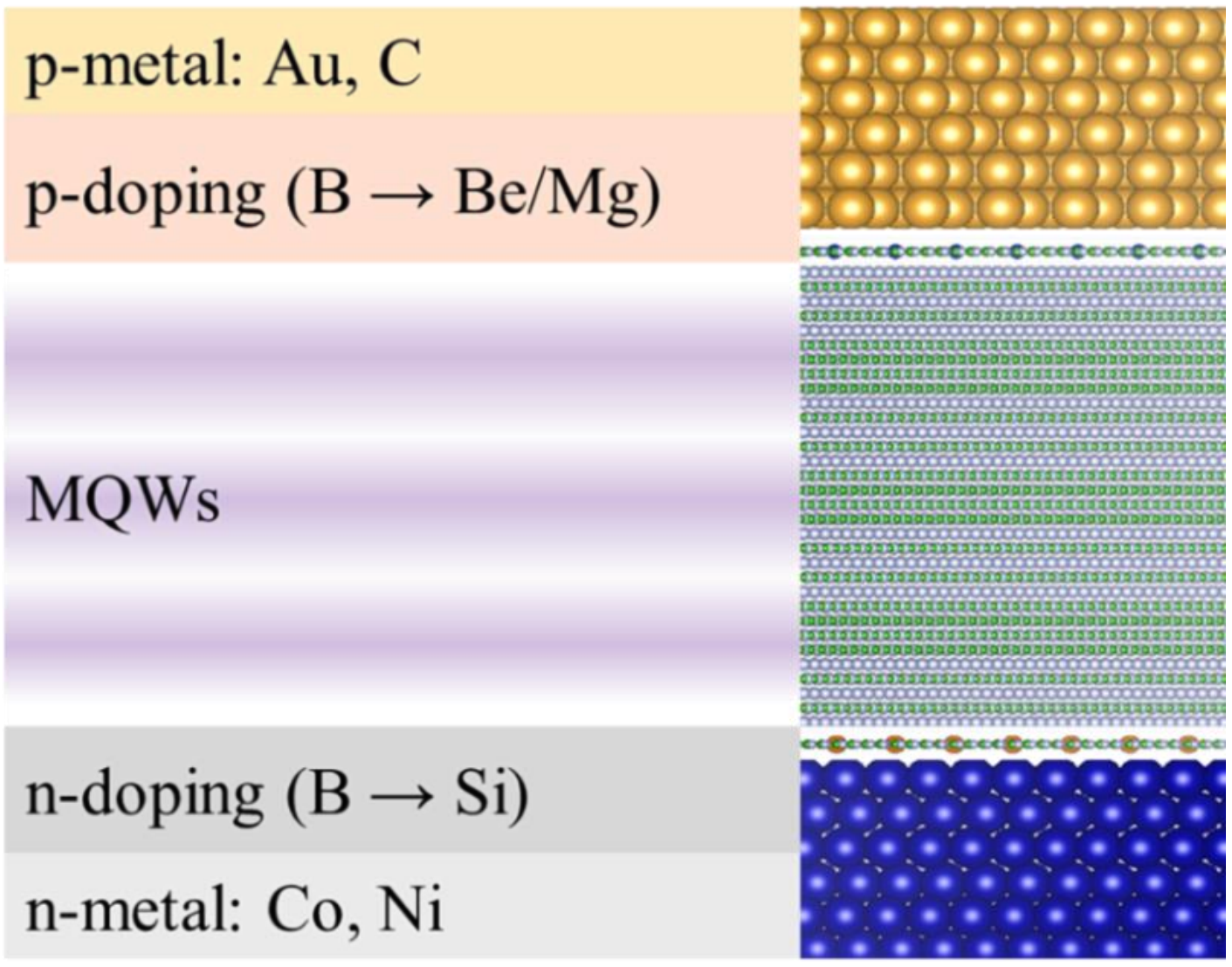

Figure 5. Design of 2D UV LED based on hBN. On the top, a thin layer of gold or graphite provides p-doping to the Be or Mg doped AA' stacked hBN. Underneath is either the MQW-T1 or MQW-T2 multi-quantum well structure that terminates on both ends with intrinsic AA' stacked hBN. Below the quantum wells is the Si-doped AA' stacked hBN (n-doping) and either a thin layer of Co or Ni.

# Supporting Information

## Figure S1 band structures of bulk hBN.

Although all the different stacking forms show indirect bandgap, they have very different electronic structures (Fig. S1). The bandgap for AA’, AB’, A’B, AB, and AA stacking forms is 4.2 (5.6), 3.6 (4.9), 3.2 (4.4) , 4.4 (5.7) and 3.1 (4.3) eV, respectively in GGA-PBE (HSE06) calculations. Moreover, the gap of each form locates in different k points.

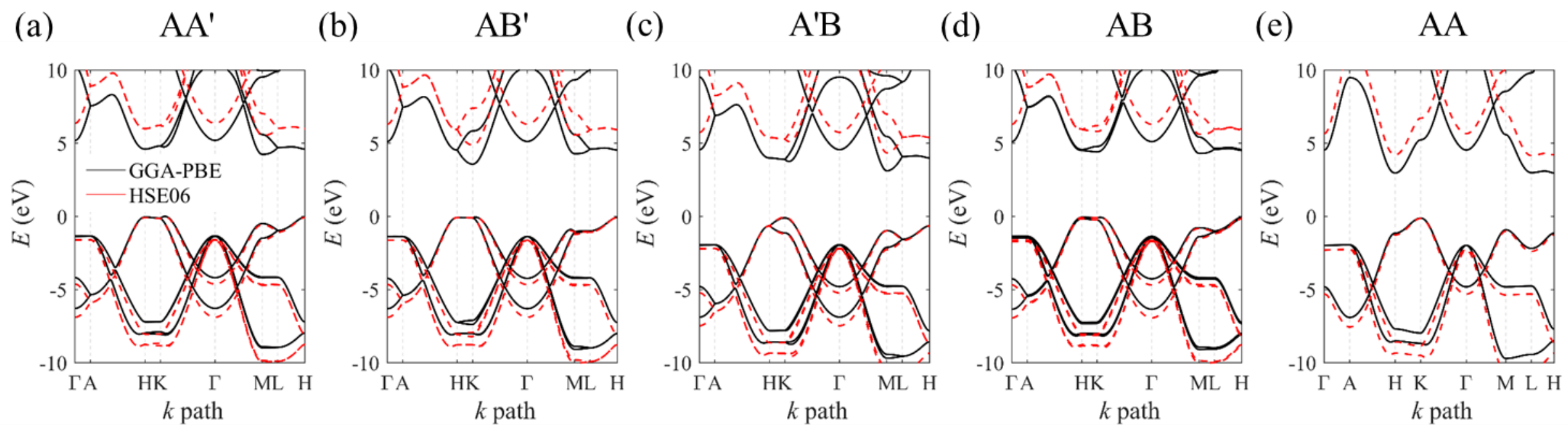

Figure S1. Band structures of bulk hBN with (a) AA’, (b) AB’, (c) A’B, (d) AB, and (e) AA stacking orders.

## Figure S2 hBN/hcp-Co(0001) and hBN/fcc-Co(111) metal contacts

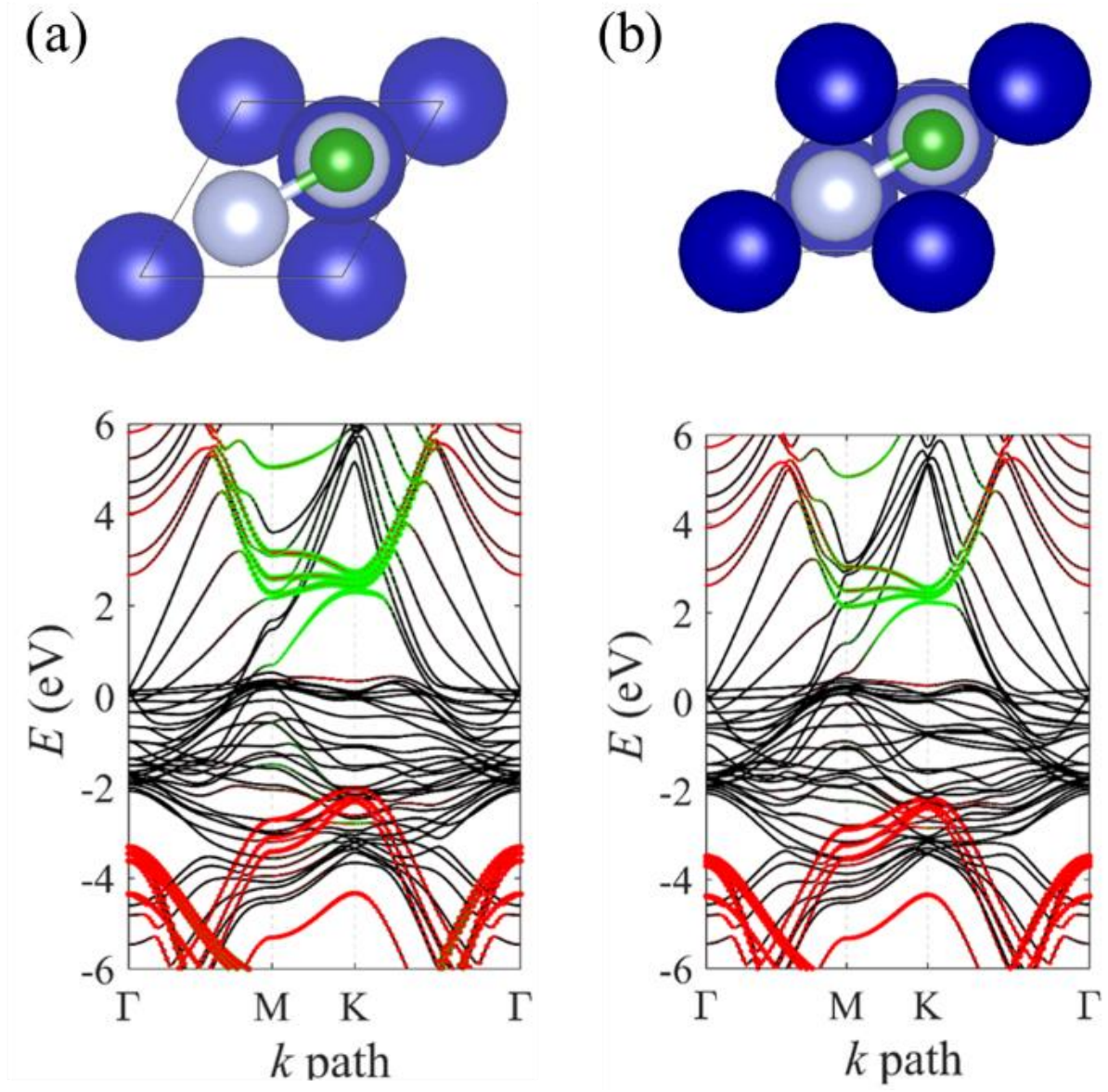

Figure S2. Atomic and band structures of monolayer hBN on a) hcp-Co(0001) and (b) fcc-Co(111). The full bands are solid lines in black, projections on B atoms are marked in green and projections on N atoms in red.

## Figure S3 doping effect on the Schottky barrier of hBN/metal contacts

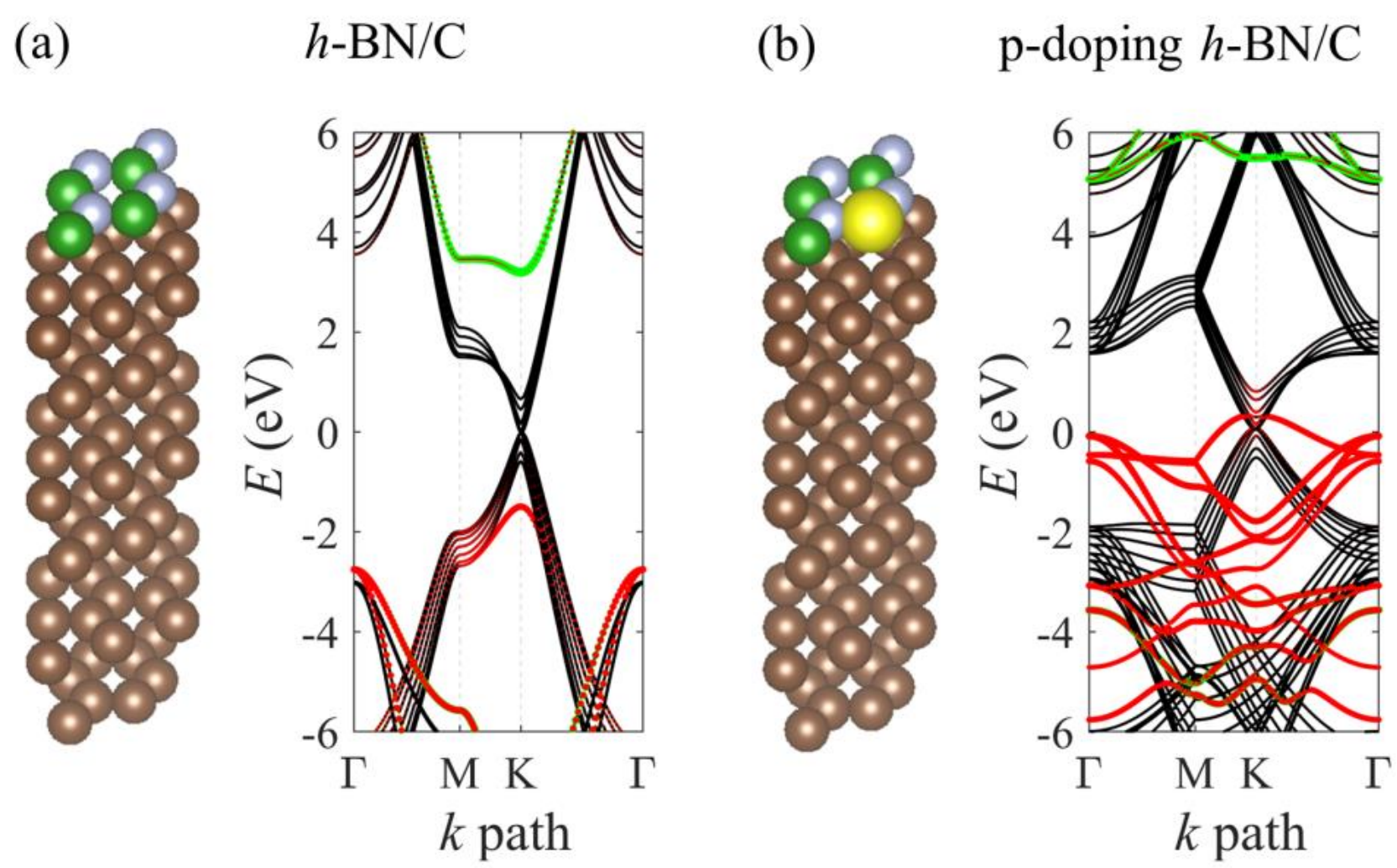

Figure S3. Atomic and band structures of (a) an intrinsic-hBN/graphite interface and (b) a 2x2 hBN/graphite interface with one B atom replaced with Bi. The full bands are solid lines in black, projections on B atoms are marked in green and projections on N atoms in red. The VASP solution of the density of states, Fermi levels and band edges result in an effective doping density of $1.4 \times 10^{22}$ $cm^{-3}$ under fully ionization assumption.